# Polarization Effects in Reflecting Coronagraphs for White Light Applications in Astronomy

**James B. Breckinridge**
Jet Propulsion Laboratory
California Institute of Technology
4800 Oak Grove Drive
Pasadena, CA. 91109
jbreckin@jpl.nasa.gov

**Ben R. Oppenheimer**
American Museum of Natural History
Department of Astrophysics
79th Street at Central Park West
New York City, NY 10024
bro@amnh.org

## Abstract

The properties of metal thin films have been largely overlooked in discussions of the technical limitations and problems that arise in the field of direct detection of exoplanets. Here, polarization properties and anisotropy properties of highly reflecting thin metal films are examined within the context of the requirements for the ultra-low scattered-light system performance of coronagraphs applied to space and ground-based high-contrast, white-light astronomy. Wavelength-dependent optical constants for highly reflecting thin metal films, taken from the literature are used to calculate the polarization-dependent transmissivity of a typical coronagraph. The effects of degraded performance on the astronomical science are examined. Suggestions are made for future work.

## Introduction

The detection and characterization of planets that orbit stars other than our sun (so called "exoplanets" or "extrasolar planets") will profoundly improve understanding of the evolution of the Universe and may lead to our discovery that mankind is not alone in the Universe. Direct detection of exoplanets, however, poses a difficult problem because of the huge contrast between the planet and the star it orbits.

The canonical contrast between an exoplanet and its parent star is often quoted as 1 part in $10^8$ for Jupiter analogs, and 1 part in $10^{10}$ for Earth analogs where the planet is small fractions of an arc second away from the parent star (e.g. Woolf and Angel 1998, and references therein). However, detailed spectroscopic study of Earth or Jupiter analogs actually requires measurement and fitting of deep absorption features, particularly if a case is to be made for biological activity on such planets. Many of these features are 100 times fainter at their depths than the broadband brightness predicted for these planets (e.g. des Marais et al. 2003). Measuring such absorption features does not depend solely on the number of photons received at the absorption feature's primary wavelength, but also on the contrast between the feature and the background starlight. Thus, we maintain that a valid target contrast for exoplanetary earth science is $10^{12}$. This ensures measurement and detailed study of these planets, not only direct detection.

Not only must such contrast be achieved, but also it must be maintained over a broad wavelength range. For evolved planets, the faint light of the companion planet shines in the thermal white-light that originates at the parent star and reflects from the planet. This reflected thermal radiation must be spatially filtered from the directly detected light of the star to enable such extreme contrasts in images. For proper collection and analysis of the planet's light, the exoplanet imaging system must have excellent broadband "white-light" performance, including both contrast and resolution.

Currently two optical/infrared instrument system architectures are being studied for NASA's Terrestrial Planet Finder mission (TPF), whose primary goal is direct imaging of an Earth-like planet orbiting a nearby star. One option is an interferometer; the other is a coronagraph. The interferometer option is based in part on technology and results from the Space Interferometry Mission (SIM). SIM is a white-light interferometer with a very high precision, laser metrology system that relies upon bore-sight nulling to achieve the needed contrast and large baselines for the resolution. The coronagraph approach may utilize a nulling technique or a mask to obscure the extra-solar system's bright central star, thus enabling direct imaging and spectroscopy of the planet (see TPF Mission Architecture studies, JPL website, planetquest.jpl.nasa.gov).

No matter which architecture is chosen, many technical details must be addressed before TPF can be constructed with confidence that it will work. Extremely high-fidelity wavefront propagation is required, with the optical system having only minuscule residual wavefront errors. Abundant research addresses wavefront control (such as through adaptive optical techniques, widely recognized as a critical component of TPF) and precisely polished surfaces. The relevant work on polished optical surfaces has largely focused on hyper polishing the substrate for a given optic, which is then coated with a standard metal thin film, such as aluminum, gold or silver to provide a surface with high specular reflectivity for broad-band white light. The substrates may be polished to an incredible accuracy (such as an RMS figure error of better than $\lambda/500$), but then, ironically, a coating, typically 0.2 to 2.0 µm thick, is deposited. These coatings are hundreds of times thicker than the smallest feature on the surface to be coated, and little attention is paid to the homogeneity of the coating. (The authors note, however, that optics built for high-energy photons such as X-rays are often coated with extreme care, e.g. Windt and Waskiewicz 1994. This care is typically limited to optics smaller than 20 cm in diameter, due to the deposition technique. Furthermore, the coatings have bandwidths limited to less than a few percent at best.) In fact, the properties of these coatings have a significant effect on the light propagated through the system—significant at least in the case where the point spread function (PSF) or the wavefront error at the science detector needs to be understood at the level of 1 part in $10^8$ or $10^{12}$. This paper is designed to make some of those effects apparent to optical/infrared astronomical instrument builders, and to incite new research into these issues. We note that some of these effects persist even in the case of an utterly perfect coating with no inhomogeneities. We show that such effects must be accounted for in systems where a contrast of one part in $10^8$ or $10^{12}$ is required.

Our calculations mainly concern reflective coronagraph designs, but many of these calculations are applicable or extensible to interferometric systems. We concern ourselves only with reflective systems, because with current technology, refractive systems seem impractical. For instance, planned coronagraphs require telescope apertures in excess of 2, probably 4, meters in diameter. The largest transmission optical element in scientific use in the world is the Yerkes 1-meter (40-inch), at least two times smaller than the elements needed for coronagraphs. No demonstration has succeeded in building refractive optical elements larger than approximately 1.2 meters that are structurally supported in the center properly without enormous chromatic aberrations. One might imagine a very large transmissive element in space where there are no effects from gravity to sag the lens, but even partial correction of the longitudinal and lateral chromatic aberration over the required optical bandwidth would necessitate many large transmissive elements in the design. The areal density of a refractive optical element is so high that apertures in excess of 2 meters are prohibitively expensive in terms of throughput and money. The primary optical element for a coronagraph will most probably be a reflecting mirror consisting of both the substrate that is figured and polished, and a highly reflecting metal optical thin film that is supported by the substrate to provide the required broadband reflectivity. In writing this paper we found that significant work on metal, thin films for astronomy is lacking but required before TPF sees the light of day from another world.

In particular, this paper explores, as an example, the significance of the polarization properties of metal, thin films used in astronomical optical systems. Polarization effects contribute to the quality of the final image. In this case, the polarization is due to the properties of an isotropic highly reflecting thin film, and to inhomogeneities within the thin films.

After reviewing some literature, we examine the case of a perfect (electronically isotropic) thin film on a perfectly figured optical substrate. We show that variations in the polarization reflectivity of these thin films across the reflecting surface of the primary mirror act to apodize the beam. This effect reduces the angular resolution of the imaging system. We show that even in a perfectly manufactured reflective optical system, the polarizing properties of the reflective thin films limits the angular resolution to worse than the system's diffraction limit. Second, we examine the effects of inhomogeneities in the thin films, as they exist in real systems. This effect results in a background of scattered light, an additional source of noise in the problem of detecting exoplanets. A third subject, beyond the scope of this short paper, is the fact that the integrated optical path through the imaging system is slightly different for each polarization vector. We discuss this briefly, in an effort to encourage further research, which is beyond the scope of the labs either author runs or can use.

**Prior Research on Astronomical Mirror Coatings and Polarization**

Hiltner and Schild (1965) and Appenzeller (1966) made the observation that the best mirrors available at the time polarized incident light by approximately 6%. That is, if one were very careful to have "totally" unpolarized light at normal incidence on a concave astronomical mirror the light reflecting off at normal exitence would be 6% linearly polarized.

Breckinridge (1971) measured the effects of polarization on astronomical imaging spectrometer data for solar applications and found these effects to have a profound effect on the quality of the data.

Born and Wolf (1980) showed that if light is incident on a reflecting mirror at an angle, the reflectivity of light polarized in the plane of incidence does not equal the reflectivity of light polarized perpendicular to the plane of incidence. The steeper the angle (up to about 60º), the more the reflectivities differ. The real and imaginary parts of the complex index of refraction characteristic of metals are wavelength dependent. Therefore the polarization content of the reflected beam is wavelength dependent.

Breckinridge, Kuper and Shack (1984) analyzed the properties of an all-reflecting coronagraph and discussed manufacturing tolerances required to control scattered light to the levels required for planet detection. Ward (1988) provides a summary of the optical methods for determining the optical constants of bulk materials and films and shows how these methods developed into the common practice of ellipsometry used by metallurgists.

Chipman (1987) provided a wavefront theory analysis tool for modeling the end-to-end system polarization transmissivity as a function of FOV. He wrote a ray-trace program for end-to-end modeling of the polarization properties of optical wavefronts within reflecting and refracting broadband, "white-light" remote sensing instruments and telescopes. An earlier version of this modeling tool was used by Breckinridge and Schindler (1981) to ray trace the polarization properties of Fourier Transform Spectrometers.

More recently, optical engineers have used specially designed stacks of dielectric thin films to control the polarization properties of thin metal reflecting coatings. MacLoed (2001) shows that high reflectivity for all polarizations is achievable for relatively narrow optical bandwidths, of a few hundreds of angstroms.

This series of papers provided the heritage upon which the current paper is based and strongly suggested that some relatively simple effects warranted investigation in terms of extremely high-contrast imaging, such as that needed by TPF.

**Perfect, Ideal Metal-Coated Mirrors**

Telescope primary mirrors are curved concave to concentrate the very faint starlight onto relay and analysis optics and to form a high quality image of object space at the focal plane. A ray from the object, which reflects from the edge (marginal ray) of the mirror to the focus, has reflected off the mirror at an angle larger than those rays closer to the vertex of the mirror. As the F-number decreases, the marginal ray angle deviation increases, where the marginal ray angle is simply arctan $(0.5F^{-1})$, where $F$ is the F-number of the imaging system.

We examine the expected polarization from a perfectly deposited thin film on an optical substrate as a function of the angle of reflection. We calculate of the reflectivity in both s and p polarizations of light incident on ideal aluminum, silver and gold metal films as a function of angle of incidence and of wavelength, both using Maxwell's equations (Born and Wolf, 1980) and verified using a commercial ray trace design program provided by Professor A. MacLeod (2002) and implemented by N. Raouf, (2003). In both cases the known indices of refraction for each of the three evaporated thin metal films (Palik 1985, page 373), each as a function of wavelength, were used for the calculations. The real and imaginary parts of the index of refraction of the these thin metal films vary by ~1%. This variation depends on deposition techniques, as discussed by Palik (1985), but first we consider only an ideal thin film.

Here we provide the tools to estimate the reflectance of a metal mirror as a function of linear polarization. Based on measured properties (real and imaginary parts of the index of refraction) the polarization reflectivities for the electric vector perpendicular to the plane of incidence (the transverse electric wave), which we will write: $R_{te}$, and its orthogonal companion the electric vector parallel to the plane of incidence (the transverse magnetic), which we will write: $R_{tm}$, are calculated. To remind the reader, the plane of incidence is the plane defined by the vectors of propagation prior to and after reflection (or refraction). After reflection from a metal mirror tilted at an angle, $\theta_1$ the radiation is no longer unpolarized, or rather $R_{te} \neq R_{tm}$.

Reflecting mirrors on astronomical telescopes are often processed to be thin to assure good adherence to the substrate and to assure that the shape of the reflecting surface precisely conforms to the optically figured and polished substrate. A dielectric coating often covers the highly reflecting metal mirror for two purposes: 1. Mechanical protection of the surface, for example, to inhibit oxidation and 2. Enhancement of the reflectivity over a broad bandwidth. The dielectric over coating does not play a major role in the polarization properties of the highly reflecting conducting mirror designed and built for remote sensing astronomical applications and therefore we shall neglect it at this time.

We will use the notation applied by Born and Wolf (1980), chapter 13 section 4 to derive equations for the reflectivities and then convert the real and imaginary indices of refraction to the standards used in modern materials tables. It is important to note, to avoid additional confusion on this subject, that there are several ways in use today to express the index of refraction of a thin film metal conductor. Born and Wolf follow the German tradition and express the complex refractive index of the film, $\hat{n}$ as given by $\hat{n} = n(1 + jk)$, whereas the French tradition is to express the complex refractive index of the film, $\hat{n}$ as given by $\hat{n} = n - jk$.

Today most materials scientists who apply the polarization properties of thin films to study materials using methods such as ellipsometry use the form $\hat{n} = n - jk$, whereas theoretical physicists such as Wolf use the form $\hat{n} = n(1 + jk)$. For example the tables by Palik (1985) use the form $\hat{n} = n - jk$. We stick to the Born and Wolf notation in the derivations of the reflectivity equations below.

Consider a quasi-monochromatic unpolarized plane wave passing within a dielectric medium of index $n_1$ and incident onto a highly reflecting, but absorbing, optical thin film (conductor) at angle $\theta_1$. The conducting metal optical thin film is of thickness h. A portion of the wavefront propagates into the metal whose index of refraction $\hat{n}_2 = n_2(1.+ jk_2)$, where $n_2$ and $k_2$ are the real and imaginary parts of the index of refraction. The angle between the wavefront that propagates within the conducting medium and the normal incidence is $\theta_2$. Born and Wolf (1980) treat the case of a partially transparent conducting substrate. This quasi-monochromatic wavefront passes through the absorbing substrate of thickness h, and exits to pass into another dielectric (the substrate) of index $n_3$ and propagates at angle $\theta_3$.

Since most astronomical mirror coatings are designed to be opaque we will assume that the coating system sees no reflection from the surface at the interface between medium 2 (the conductor) and medium 3 (the dielectric substrate figured mirror). This assumption is equivalent to specifying that the light does not interact with the glass (or other) substrate.

Born and Wolf (1980), section 1.6.4, equations 55 show that, for a wavefront traveling in a dielectric of index $n_1$, incident on dielectric $n_2$, the amplitude reflectivity for the electric vector perpendicular to the plane of incidence, $[r_{12}]_{te}$, is

$$[r_{12}]_{te} = \frac{n_1 \cos\theta_1 - n_2 \cos\theta_2}{n_1 \cos\theta_1 + n_2 \cos\theta_2} \ . \tag{1}$$

The intensity reflectivity for either the electric vector perpendicular to the plane of incidence or parallel to the plane of incidence is given by $R_{12} = |r_{12}|^2$.

Born and Wolf (1980) further show that the formulae relating to the reflection and transmission of a plane quasi-monochromatic wave are obtained by replacing $n_2$ in equation 1 by $\hat{n}_2 = n_2(1 + jk_2)$. We will find that it is convenient to set

$$\hat{n}_2 \cos\theta_2 = u_2 + jv_2, \tag{2}$$

Where $u_2$ and $v_2$ are real and introduced to simplify the algebra. It is straightforward to express $u_2$ and $v_2$ in terms of the angle of incidence and the constants that characterize the optical properties of the first and second medium. Born and Wolf (1980) follow in the tradition of Abeles (1963) and Abeles (1957) to square equation 1 and use the law of refraction, $\hat{n}_2 \sin\theta_2 = n_1 \sin\theta_1$, to obtain:

$$(u_2 + jv_2)^2 = \hat{n}_2^2 - n_1^2 \sin^2\theta_1 \ , \tag{3}$$

where Born and Wolf (1980) use $\hat{n}_2 = n_2(1 + jk_2)$.

On equating the real and imaginary parts, we obtain the system of equations needed to solve explicitly for $u_2$ and for $v_2$:

$$u_2^2 - v_2^2 = n_2^2(1 - k_2^2) - n_1^2 \sin^2\theta_1 \tag{4}$$

and

$$u_2 v_2 = n_2^2 k_2 \tag{5}$$

Equations 4 and 5 are two coupled equations. Methods of numerical analysis are the most appropriate and efficient computational tools to employ to solve explicitly for $u_2$ and for $v_2$.

Born and Wolf (1980) solve on page 628, for the amplitude of the reflectance $\rho_{12}$ and for the phase change $\phi_{12}$, for the electric vector perpendicular to the plane of incidence (the TE wave). As noted above the intensity reflectance is $|\rho_{12}^2|$. Thus,

$$R_{te} = |\rho_{12}^2| = \left| \frac{(n_1 \cos\theta_1 - u_2)^2 + v_2^2}{(n_1 \cos\theta_1 + u_2)^2 + v_2^2} \right|, \tag{6}$$

$$\tan\phi_{12} = \frac{2 v_2 n_1 \cos\theta_1}{u_2^2 + v_2^2 - n_1^2 \cos^2\theta_1} \tag{7}$$

Born and Wolf (1980) solve, on page 629, for the amplitude of the reflectance $\rho_{12}$ and for the phase change $\phi_{12}$, for the electric vector parallel to the plane of incidence (the TM wave) and obtain:

$$R_{tm} = \left|\rho_{12}^2\right| = \left|\frac{\left[n_2^2(1-k_2^2)\cos\theta_1 - n_1 u_2\right]^2 + \left[2n_2^2 k_2 \cos\theta_1 - n_1 v_2\right]^2}{\left[n_2^2(1-k_2^2)\cos\theta_1 + n_1 u_2\right]^2 + \left[2n_2^2 k_2 \cos\theta_1 + n_1 v_2\right]^2}\right|, \quad (8)$$

$$\tan\phi_{12} = 2n_1 n_2^2 \cos\theta_1 \left|\frac{2k_2 u_2 - (1-k_2^2)v_2}{n_2^4(1+k_2^2)^2 \cos^2\theta_1 - n_1^2\left(u_2^2 + v_2^2\right)}\right|, \quad (9)$$

Therefore it is straightforward to calculate intensity reflectance for the electric vector perpendicular to the plane of incidence (using equation 6) and for the electric vector parallel to the plane of incidence (the TM wave), using equation 8.

This analysis treatment is provided to the reader to give background to the physics (condensed matter) and the mathematical methods used today in studies of the interaction of light and metal thin films. The polarization reflectivity numbers used in this paper were generated with the most advanced computer-aided synthesis and design program used in commercial applications (MacLeod 2001, 2002).

$R_{tm}$ and $R_{te}$ were calculated for three different metal mirror surfaces, gold, silver, and aluminum, at angle of incidences ranging from 0 to 30º using the MacLeod (2002, 2001) software operated by the professional thin film designer N. Raouf (2003). We calculate the case where the dielectric medium is vacuum, and therefore $n_1 = 1.0$, in the above equations. The indices of refraction for the thin films are taken from Palik (1985) and are a function of wavelength.

The results of the calculation are given in Figures 1, 2 and 3, where the percentage of polarization $(R_{tm} - R_{te})/(R_{tm} + R_{te})$ is plotted for various wavelengths between 300 and 1100 nm as a function of angle of incidence of the beam. The fact that there is any structure at all in the curves shown in Figures 1 to 3 means that the s ($R_{te}$) and p ($R_{tm}$) reflectivities differ. They differ increasingly as a function of the angle of incidence, and not necessarily monotonically as a function of wavelength.

What is the significance of this? The values of $R_{te}$ and $R_{tm}$ are well above 98% in most cases, but they are a strong function of the angle of incidence. This means that in a telescope with a concave reflective primary, the incident starlight will be reflected at varying angles as a function of the distance of the incoming ray from the optical axis. We can simply calculate the angle that the marginal ray in a telescope beam makes with the primary mirror as a function of the F-number, as we mentioned above. If $F = 1.5$, a suggested value for a TPF design, the marginal ray experiences a 10º angle of incidence to the metal thin film on the primary mirror. Combining the angles of incidence over the entire telescope beam with the polarization calculations for a given coating results in an apodization function being imposed on the telescope beam. This apodization function is greatest at the beam edge (the marginal ray). The apodization tapers to zero at the precise center of the beam in an on-axis telescope design. In the case of an optimal off-axis design, which is not obscured by the secondary mirror support structure (e.g. Tokunaga et al. 2002), the apodization function will not reach zero anywhere in the aperture. An example of an apodization function for an off-axis primary mirror is given in Figure 4. The apodization function is also a function of wavelength. It is important to note that in order to reduce the effect of this apodization, telescopes must be designed with long focal lengths (large F-numbers).

The detailed shape of the white-light PSF from a telescope whose primary mirror is a perfectly smooth thin film metalized mirror will differ from the well-known $[2J_1(x)/x]^2$ Airy pattern given by scalar diffraction theory. Radiation that forms the narrow core of the PSF, and thus establishes the system resolution, reflects from the edge, or the steepest portion of the mirror of the mirror. This light has the greatest difference in polarization content, and thus the greatest attenuation. Since light orthogonally polarized does not interfere and thus contribute to image formation, radiation from the edge of the mirror will not combine with the radiation from the center area of the mirror to form the "classical" $[2J_1(x)/x]^2$ Airy pattern characteristic of the theoretical diffraction limit. Figure 4 shows the apodization function for an F/1.5, off-axis, primary mirror coated with a perfect silver film and used in an imaging mode. The mirror is assumed to be a section of a parabola one half its diameter off-axis. The apodization function shows curvature around the axis of the parabola and depresses the throughput by about 1% on the outer edge (left) of the pupil. Figure 5 is the difference between PSFs made from an unapodized circular aperture of the same diameter shown in Figure 4 and the apodized aperture shown in Figure 4. A quantitative assessment of Figure 5 is shown in Figure 6, which shows vertical and horizontal slices through the image in Figure 5. The PSF is affected at the level of 7 parts in $10^5$ at angles greater than $\lambda/D$, where $\lambda$ is the observing wavelength and $D$ is the telescope diameter. Because many novel coronagraphic techniques rely on masks close to the $\lambda/D$ size (e.g. Soummer et al. 2003), and because image suppression at the level of $10^{10}$ to $10^{12}$ is required for TPF science goals, this effect must be accounted for in any TPF design.

In qualitative terms the apodization effect is diffractive. It results in a wider diffraction pattern for the telescope, by as much as 10% ($1.34\lambda/D$ instead of $1.22\lambda/D$ for the maximum radius of the now elliptical, first null in the pattern) depending on the metal coating. (The worst case is for aluminum.) In general this is not of much concern. An increase in the width of the point-spread function's (PSF) core by a few percent is generally not important for many science applications, although the $1.22\lambda/D$ rule of thumb for the angular resolving power of a telescope clearly is not correct when one considers this effect. For exoplanet imaging, where angular resolution is important, we strongly maintain that this effect must be accounted for in system designs. Indeed, strong arguments can be made that the majority of Earth-like planets orbiting nearby stars will be situated within an angle a few times $\lambda/D$, even in the most ambitious TPF designs.

The other effect of wavelength-dependent apodization is that the rings in the perfect Bessel function are reduced in significance, essentially blurred. The same effect applies to the residual speckles in the final image. These two effects are actually beneficial to the exoplanet detection problem. Indeed, some novel techniques for high contrast imaging utilize strongly apodized entrance pupils to smooth out the background of light in the PSF due to residual wavefront errors. See, for example Soummer et al. (2003), who suggests that a heavily apodized entrance pupil can substantially improve the detection limits of a standard stellar coronagraph.

One of the interesting implications of this study is that the diffraction limit of telescopes is not only limited by the telescope aperture, but also the reflective properties of the metal coating used on the mirrors in the imaging system. This is a small, but important effect, but to our knowledge, it has not been pointed out before in the literature. Taking this at face value, most astronomical instruments designed to operate at the "Nyquist Frequency" in the image plane are actually marginally over sampling the PSF. However, we would also like to note, although this is a minor digression from the purpose of this article, that Nyquist sampling of the image plane is likely insufficient for extremely high contrast imaging problems. The modulation transfer function (MTF) of any imaging system exhibits non-zero power at frequencies that exceed the Nyquist frequency. In applications where two point sources of relatively similar brightness are to be discerned, this has no importance whatsoever. However, if the PSF is to be removed at a level of one part in $10^{12}$, a residual MTF of one part in $10^6$ (typical for today's telescope instruments) beyond the Nyquist frequency may pose a significant challenge. From this point of view, the smaller the pixels the better understood the PSF will be. Of course an infinite number of pixels cannot be supplied, and actual system design will have to consider these competing factors. The fact that the MTF exhibits some power at frequencies above the Nyquist frequency has been discussed in the optics field. See, for example, Shaw and Dainty (1974) or Goodman (1970) or Reynolds, et. al. (1989). The effect is largely due to the fact that the image formation uses polychromatic, partially coherent radiation.

In addition to the polarization effects, since both the real part and the imaginary parts of the index of refraction of metals are wavelength dependent, the image formation equation in white light has a chromatic term, further corrupting the ideal point-spread function. In other words, the PSF will not be achromatic, even in the case where an ideal, perfectly deposited metal coating is applied to a perfect optical surface.

Finally, we note (with thanks to S. Shaklan for bringing this to our attention) that the s and p polarization vectors will encounter the curved optical surfaces at slightly different optical path lengths (equations 7 and 9 above). This is due primarily to the geometry of the beam and its polarization vectors. A wave with only the p polarization state incident on a perfectly coated off-axis primary mirror will emerge with an astigmatism term as compared to a circularly symmetric system. At the same time, a wave with only the s polarization will have an astigmatism term rotated 90º from the p state. A standard adaptive optics system cannot simultaneously correct both states. Thus a low-order wave front error will propagate through the coronagraph. It is important to note that in the case of TPF contrasts and science requirements, a fraction of an Angstrom of wave front error can affect the contrast by factors of 10 to 100 close to the center of the image, where, we noted above, the majority of Earth-like planets are believed to exist. This effect is wavelength dependent as well.

In practice this can be removed by separating the two polarization states and invoking an adaptive correction (through a standard deformable element) separately, with a separate wavefront sensor for each polarization state. (Fortunately some work on this issue is already in progress by Dr. Shaklan's group at JPL.)

One might think that simply inserting polarizing filters in appropriate places in the optical system might negate the effects described above, while sacrificing some throughput. However, this is only true if the film were perfectly isotropic in its properties. As we discuss in the next section, the film is characterized by significant microstructure: randomly oriented clusters of metallic conducting crystals. At the levels of control required for terrestrial planet searching, the metal surface will not perform as a homogeneous conductor, but rather there will be circular polarization appearing as the "cross-product" terms within a slightly anisotropic metal. That will be very hard to deal with, and may become the dominant source of scattered light at the very low levels. We argue that work is needed to understand the electronic properties and microstructure of conducting films of these sizes and aspect ratios (200 nm thick and >4 m across).

**Imperfect Metal Thin Films**

It has been recognized for years that the scattered light performance of the large primary mirror strongly depends on the smoothness of the mirror substrate and much effort has gone into understanding the engineering processes to create a super-smooth surface. The large area-reflecting surface of these substrates, however, does not have the quality specular reflectance required for the mirror. Thus, the substrate must be coated with the highly reflecting metal thin film. If this thin metal film is perfectly isotropic then the limiting effects discussed in §3 will dominate. Unfortunately, metal thin films over the large areas required for astronomy are not, in general isotropic (amorphous). For example, a common result of modern silver deposition techniques is the development of columnar microstructure. Dirks and Leamy (1977) provide a summary of research on this microstructure in vapor deposited thin films.

A quantitative understanding of the effects of metal thin film inhomogenieties on system scattered-light is provided by analysis of the Smith-Purcell Effect is discussed by Smith and Purcell (1953) and by Haeberle et al. (1997).

An electromagnetic light wave incident on a metal surface is known (Jackson, 1999) to induce an oscillating electronic charge inside the highly conducting metal film on the mirror substrate. The period of oscillation is characteristic of the frequency of the source: $\sim 10^{14}$ Hz. If this highly conducting metal has isotropic conductance over its entire volume (~1 μm thick by four or more meters in diameter), then the images formed by the mirror will be, "perfect" within the constraints of diffraction produced by the limits of the clear aperture, and the constraints of the polarization effects introduced by the Fresnel reflection effects discussed in §3 above.

In one of the earlier studies of this, Horowitz (1983) showed that metal thin films are not isotropic. He determined that the degree of polarization introduced by a reflecting Al metal surface depended on the angle of deposition of the metal onto that surface. Horowitz also showed that the microscopic grain structure of the deposited Al metal was not isotropic and varied across the surface: being columnar in some regions and amorphous in others. This anisotropy causes the electronic conductance to vary across the surface and may cause wavefront deformations that dominate the system performance.

Hodgkinson (1988) provided a review of the micro structural induced anisotropy in optical thin films. Gee and Wilson (1986) discuss reflection anisotropies in telescope mirror coatings. Hodgkinson (1990, 1991) measured optical anisotropy in thin films deposited obliquely and made in situ observations and developed a computer model to explain his observations.

Ironically, very little work has gone into optimal deposition of metal thin films onto these super-smooth surfaces (except for small optics used in X-ray imaging devices; Windt and Waskiewicz 1994). The imperfections in the metal thin films generally manifest themselves in the form of unwanted, usually columnar, structure in the film after it has been deposited on the optical substrate surface. High spatial frequency imperfections in metal thin films result in scattering light through large angles in the optical system. If a large fraction of the mirror's area contains an anisotropic structure, through preferential polarization, or preferential conduction in certain directions in the metal coating, then unwanted radiation can be scattered into the field of view. In the case of TPF science, even very small patches of anisotropy could scatter enough light to make the background overwhelm objects $10^{10}$ times fainter than the bright star in the field of view.

Figure 7 shows an example of this columnar structure and nano-roughness in a film of zinc sulphide deposited on glass. The tiny features on the upper surface of this film (the optical surface) will act to scatter light as described above. MacLeod (2001) reports:

> "The columnar structure is common to metallic films as well as some dielectrics. I would be inclined to say that vapor deposited films tend to have a columnar structure that is most pronounced in thermal evaporation and less pronounced in the energetic processes like sputtering."

Hodgkinson (2003) attributes this to a rapid oxidation of the tiny vapor droplets as they adhere to the glass substrate. The droplets are oxidized on the edges not in contact with the glass itself, thus giving a preferential direction to the oxidation structure (away from the surface).

**Recommendation for Further Research**

Clearly additional work is needed to understand the effects of real metal thin films on coronagraphy and interferometry. Realistic computer models of the polarization-dependent point spread function need to be constructed using properties of real materials and an optical system point design in order to include the polarization effects integrated over each powered and tilted optical surface.

An estimate of the magnitude of the effect of thin-film anisotropy can be obtained by examining two images of a filled spherical aperture from the center of curvature. The mirror would be illuminated from the center of curvature in linearly polarized light in one direction. Then, the image of the filled pupil is acquired from the reflected light through an orthogonal linear polarizer with a high dynamic range CCD. The polarizers would be positioned to null the light from most of the mirror's surface, revealing only the patches on the mirror where a changed polarization state is induced by one or more of the effects discussed above. These patches will appear as brighter spots on surface of the mirror. Since this suggested experiment illuminates the mirror from the center of curvature the apodization from center to edge (§3) will not be apparent: all rays that compose the final image strike the primary at normal incidence.

Furthermore, innovative ways to measure radiation at less than one part in $10^8$ or $10^{12}$ need to be developed. Electronic imaging devices are not capable of such contrast in single images.

The highly reflective optical thin film, at the surface of dynamic adaptive optics deformable mirrors, rests on a series of small actuators, each of which moves relative to its neighbor to compensate for wavefront error. This dynamic flexing of the fragile thin film may alter the physical structure of the film, changing the thin film local anisotropy, and introducing unwanted scattered light and other, more complex polarization effects than those discussed above. Indeed, angles of incidence on the silver film coating most deformable mirrors are highly variable in space and time.

The polarization-apodization effects might be controllable using appropriately designed, and fabricated (by an as yet unknown process) masks or randomly addressable digital devices to control the complex amplitude phase and polarization transmittance to compensate for the aberrations introduced in the telescope.

If these effects limit our ability to build telescopes to detect and characterize planets around stars, a material science development program must be supported with the intention of engineering the appropriate materials and deposition techniques needed. The future of exoplanetary science depends upon a collaborative partnership between the astronomical community and the thin-film, materials science, nanotechnology community.

## Acknowledgments

One of us (JBB) wishes acknowledges the support of the National Science Foundation under the employee Research and Development grant award. The work described in this paper was partly performed at the Jet Propulsion Laboratory, California Institute of Technology, under a contract with the National Aeronautics and Space Administration. This work was also supported by the National Science Foundation under Grant AST-0215793 at the American Museum of Natural History, which, with generous support from the Museum and its patrons, established the Lyot Project for exoplanet imaging. See research.amnh.org/users/bro for additional information about the Lyot Project. We also acknowledge a rigorous reading by the referee as well as very useful comments from Stuart Shaklan and Richard Dekany, and thoroughly enjoyable discussions with Anand Sivaramakrishnan.

# Figures

Figure 1. Percent polarization induced by an aluminum coating as a function of wavelength of incident light and angle of incidence. Percent polarization is presented on a logarithmic scale to enhance the details of the surface plot.

Figure 2. Percent polarization induced by a silver coating with notation the same as in Figure 1.

Figure 3. Percent polarization induced by a gold coating with notation the same as in Figure 1.

Figure 4. Pupil apodization due to polarization effects induced by a perfectly deposited, isotropic, ideal silver thin film on a perfect parabolic surface. The surface represents an F/1.5 primary one half of the mirror diameter off-axis. The grey scale ranges from 99.9% to 98.9% reflective, with the maximum transmittance toward the parabola's axis at right.

Figure 5. Difference between an ideal PSF from a circular, unapodized, unobscured aperture and the apodized aperture shown in Figure 4. The grey scale spans $10^{-6}$ to $10^{-5}$ on a linear scale. Spatial axes indicate distance from the image center in units of $\lambda/D$.

Figure 6. Linear profiles of the residual image shown in Figure 5, to provide a quantitative analysis. The grey curve represents the image along a vertical line centered on the image and the black curve is a horizontal line, also centered on the image.

Figure 7. Electron micrograph of a region of a zinc sulphide film on a glass substrate. Part of the film has been mechanically removed, leaving the columnar structure visible in the cross-section (courtesy I. M. Reid et al. 1979).

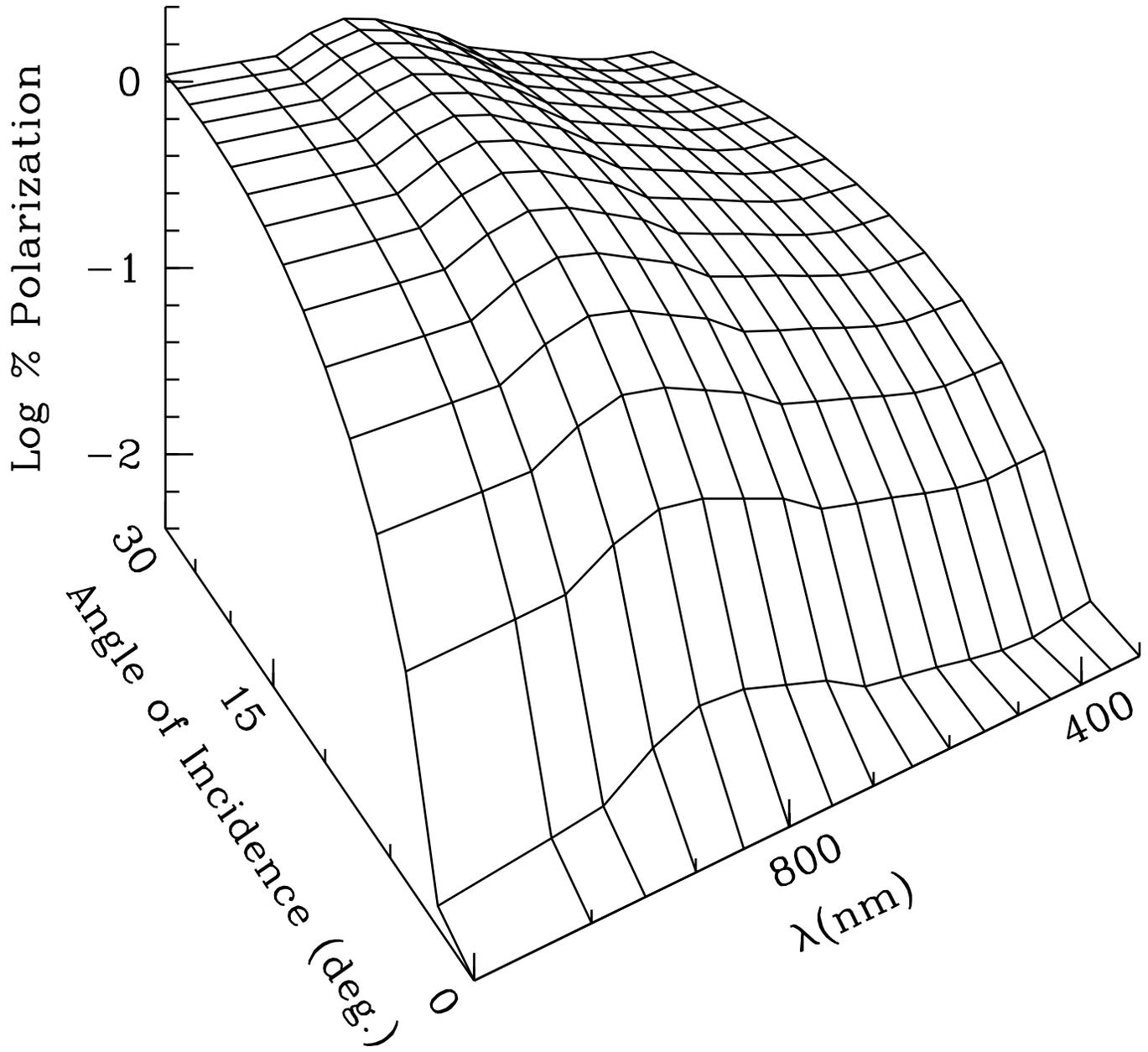

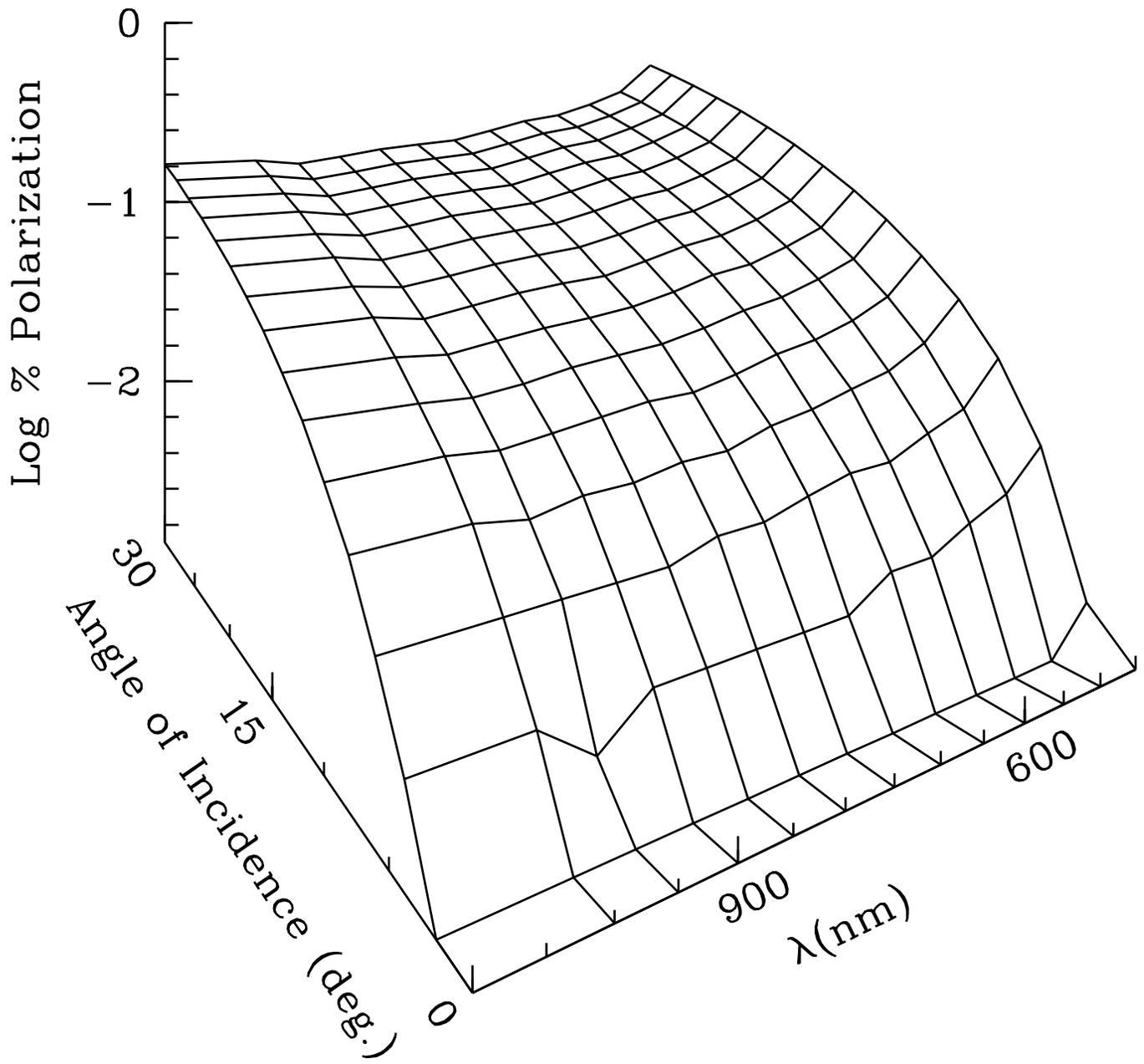

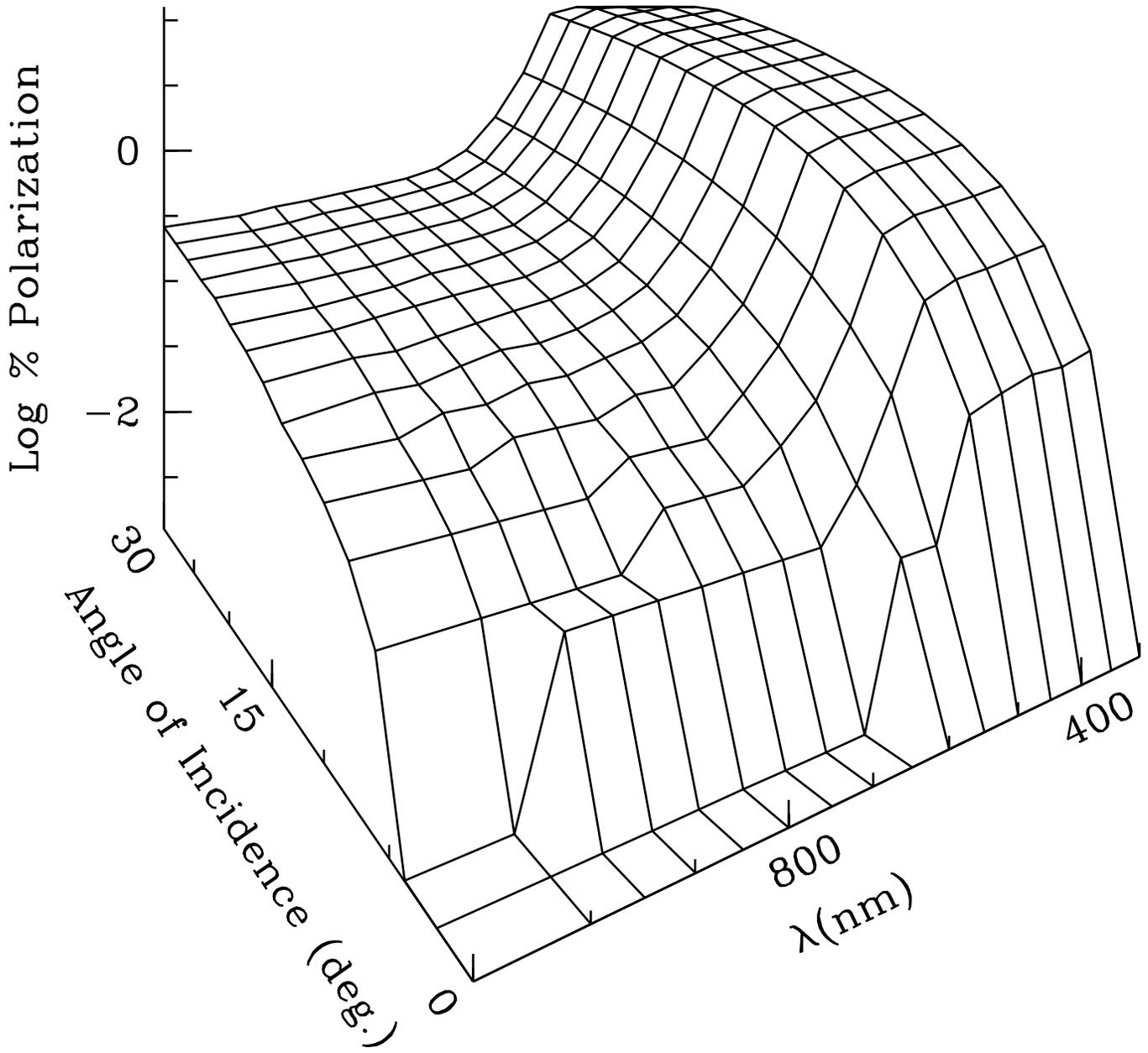

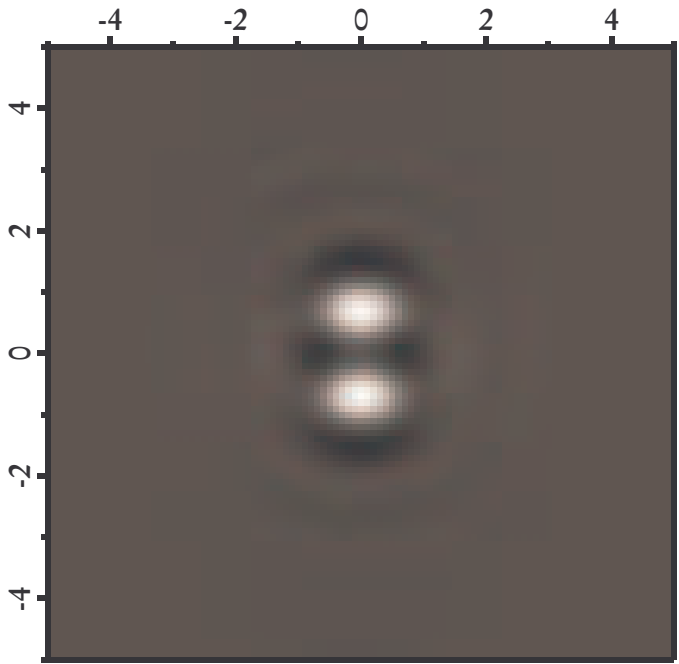

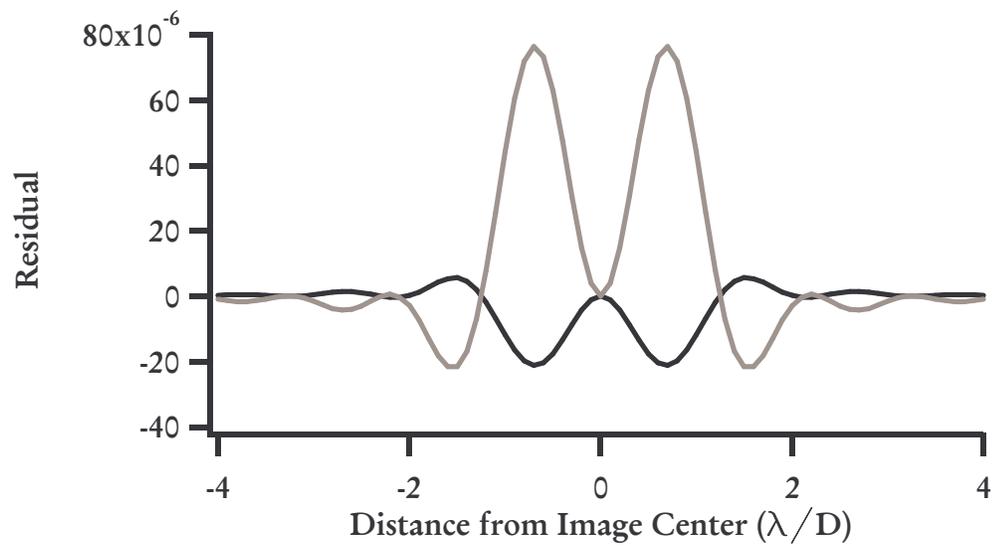

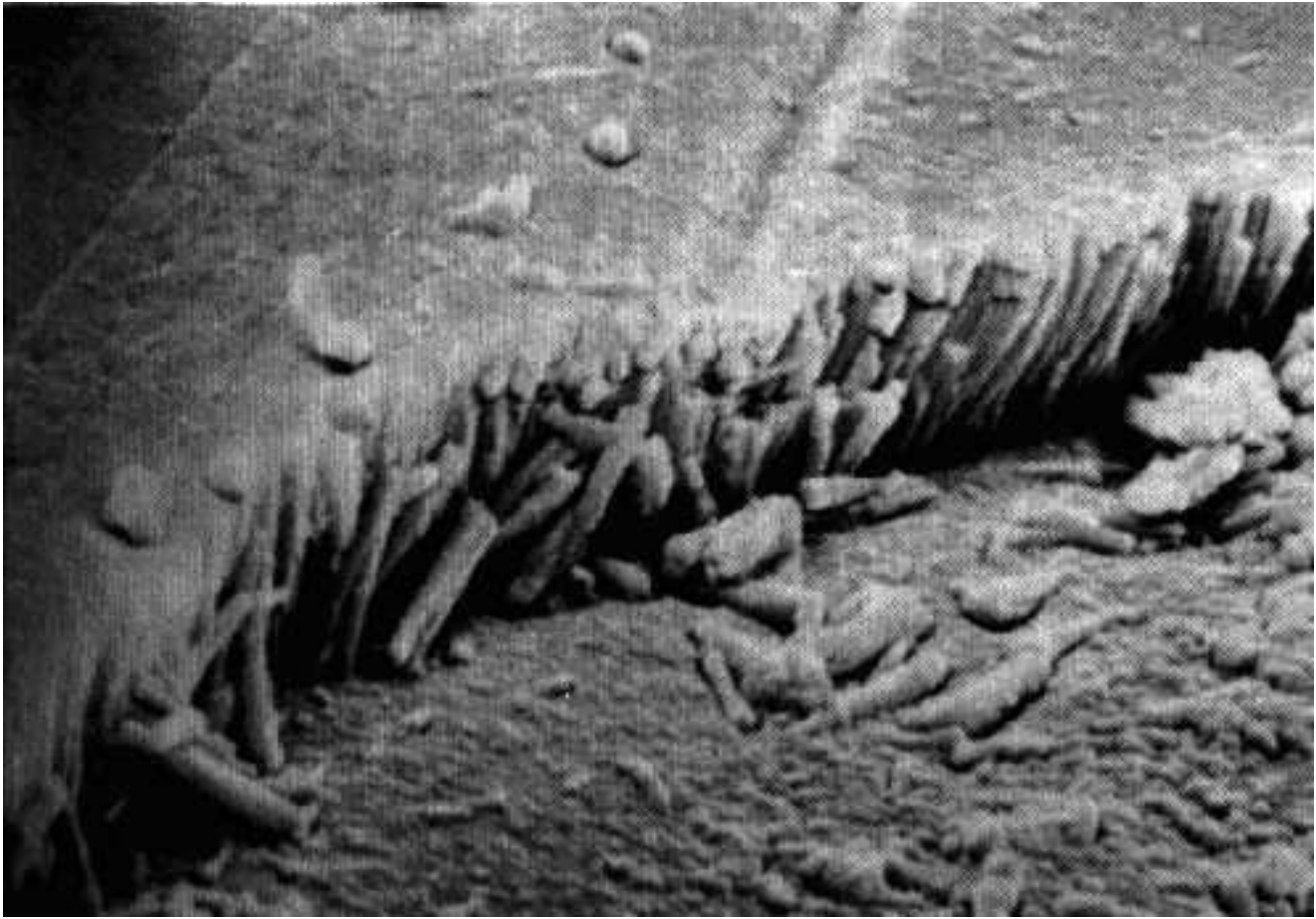